\documentclass[onecolumn,11pt,tightenlines,nobibnotes,nofootinbib,superscriptaddress]{revtex4-2}
\usepackage{amsmath,amssymb,bm}
\usepackage[a4paper,bindingoffset=0.in,left=0.65in,right=0.65in,top=1in,bottom=1in,footskip=.25in]{geometry}
\usepackage{graphicx,subfigure,epsfig}
\usepackage{bigints}
\usepackage{color}
\usepackage[breaklinks,colorlinks,urlcolor=blue,citecolor=blue,linkcolor=blue]{hyperref}
\usepackage{mciteplus}
\definecolor{lcolor}{rgb}{0.5,0,0}
\definecolor{citcolor}{rgb}{0,0.3,0.0}
\usepackage[capitalise]{cleveref}
\usepackage{todonotes}

\newcommand{\be}{\begin{equation}}
\newcommand{\ee}{\end{equation}}
\newcommand{\beq}{\begin{eqnarray}}
\newcommand{\eeq}{\end{eqnarray}}
\newcommand{\benn}{\begin{displaymath}}
\newcommand{\eenn}{\end{displaymath}}
\newcommand{\beann}{\begin{eqnarray*}}
\newcommand{\eeann}{\end{eqnarray*}}

\begin{document}

\title{Accessing the stringy structure of proton in the framework of Color Glass Condensate}
\author{Wenchang Xiang}
\email{wxiangphy@gmail.com}
\affiliation{Physics Division, Guangzhou Maritime University, Guangzhou 510725, China}
\author{Yabing Cai} 
\email{yanbingcai@mail.gufe.edu.cn} \thanks{Corresponding author} 
\affiliation{Guizhou Key Laboratory in Physics and Related Areas, Guizhou University of Finance and Economics, 
Guiyang 550025, China}
\author{Mengliang Wang}
\email{mengliang.wang@mail.gufe.edu.cn}
\affiliation{Guizhou Key Laboratory in Physics and Related Areas, Guizhou University of Finance and Economics, 
Guiyang 550025, China}
\author{Daicui Zhou}
\email{dczhou@mail.ccnu.edu.cn}
\affiliation{Key Laboratory of Quark and Lepton Physics (MOE), and Institute of Particle Physics, Central China Normal University, Wuhan 430079, China}


\begin{abstract}
To investigate the possible geometric structure of the proton, an improved stringy proton model is constructed beyond the smallest distance approximation, where the constituent quarks are connected by gluon tubes which merge at the Fermat point of the quark triangle. The exclusive diffractive vector meson production process in electron-proton deep inelastic scattering is used to test the stringy structure of the proton. We calculate the coherent and incoherent differential cross sections of the exclusive diffractive $J/\Psi$ photoproduction in the framework of Color Glass Condensate. The results show that our calculations are in good agreement with HERA data. Especially, our results give a better description of the HERA data at small $t$ as compared to the ones from the hot spot model where the constituent quarks are uncorrelated distributed in the proton. Meanwhile, the radius of the proton resulting from the improved stringy proton model is coincident with the one from fitting to the data from GlueX Collaboration at Jefferson Lab, which indicates that the predictive power of the stringy proton model is significantly improved once it goes beyond the smallest distance approximation. Moreover, we assume that the transverse shape of gluon tube satisfies Gaussian distribution, and explore the distribution width of the individual gluon tubes. We find an interesting result that the up quark induced gluon tube seems to have larger distribution width than the down quark induced gluon tube, which is favored by the HERA data.  
\end{abstract}

\maketitle

\section{Introduction}
\label{intro}
The geometric structure of proton is of fundamental interest in the field of high energy physics. It is known that the proton consists of quarks and gluons, collectively called as partons. In collider physics, it is important to understand how partons distribute within the proton, since almost all of observables require knowledge of parton distribution functions (PDFs) in quantum field theory. However, we are still far from deep understanding the behavior of partons inside the proton. Traditionally, the lepton-proton colliders are used to study the PDFs, such as HERA at DESY. A point-like particle, like electron, is used to be as a probe to resolve the internal structure of the proton, in which the electron emits a virtual photon which fluctuates into quark-antiquark ($q\bar{q}$) dipole and then interacts with the target proton. Among the lepton-proton deep inelastic scattering (DIS) processes, the exclusive diffractive vector meson production process is an extraordinary tool to probe the proton wave function at high energies (equivalently small Bjorken-$x$). 

The exclusive diffractive process requires at least two gluons to be exchanged between the $q\bar{q}$ dipole and proton target, thus the process is proportional to the squared parton density, which renders this process very sensitive to the structure of the proton. In addition, the exclusive diffractive process can provide access to obtain the transverse spatial distribution of parton in the proton at small $x$, since this process is the only process in which the transverse momentum transfer $t$ is experimentally accessible, and the $t$ is the Fourier conjugate of the impact parameter profile of the proton. Finally, the property of the exclusive diffractive scattering requires a large rapidity gap between the produced vector meson and scattered target proton due to the color singlet exchange in the diffractive DIS process. The large rapidity gap feature offers an effective way to identify the diffractive events in the experimental measurements, which then provides the cleanest data and signals to investigate the spatial structure of the proton. Consequently, the exclusive diffractive vector meson production processes not only play an important role in the HERA experiments at DESY, but also are indispensable in the future electron-ion facilities at EIC\cite{AbdulKhalek:2021gbh}, LHeC\cite{LHeC:2020van}, and EicC\cite{Anderle:2021wcy} when searching for signals of the PDFs. In this paper, we use $J/\Psi$ production in electron-proton collisions at HERA energies to study the stringy structure of the proton. 

\begin{figure}[t!]
\setlength{\unitlength}{1.5cm}
\begin{center}
\epsfig{file=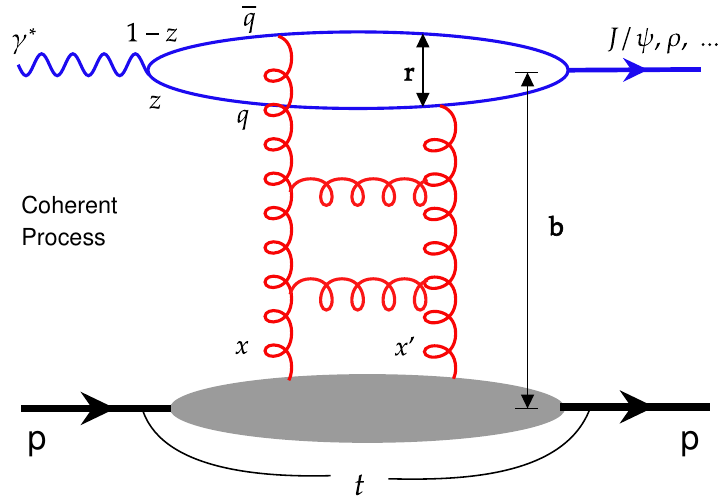, width=7cm,height=4.5cm}
\hspace{0.8cm}
\epsfig{file=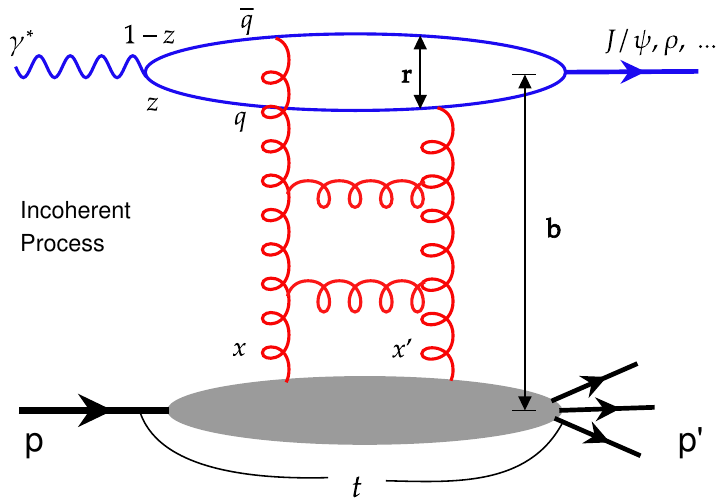, width=7cm,height=4.5cm}
\end{center}
\caption{The vector meson productions in coherent (left) and incoherent (right) $\gamma^*+p$ exclusive diffractive DIS in dipole picture.}
\label{eddis}
\end{figure}

There are two types of exclusive diffractive processes in the DIS, known as coherent and incoherent processes, see Fig.\ref{eddis}. For the coherent process, the target proton keeps intact after the scattering, the differential cross section of this process is proportional to the square of the first moment of the scattering amplitude averaged over the initial state configuration, and contains the information on the average transverse spatial structure of the target proton. In the incoherent process, the target proton breaks up after the scattering, the differential cross section of the incoherent process varies directly with variance of the scattering amplitude squared, as such probes the detailed structure of the target proton. So, we will mainly use the incoherent process to investigate the fine spatial structure of the proton in this work.        
       
There are a lot of efforts on the studies of the proton shape over the past decade\cite{Schlichting:2014ipa,Mantysaari:2016ykx,Mantysaari:2016jaz,Mantysaari:2017cni,Mantysaari:2022ffw,Mantysaari:2020lhf,Cepila:2016uku,Cepila:2023dxn,Kumar:2021zbn,Kumar:2022aly,Kumar:2024kns,Xiang:2023msj,
Demirci:2022wuy,Traini:2018hxd}. One track of these studies is based on the Color Glass Condensate (CGC) effective field theory, which was proved be to a convenient and efficient method to calculate the quantities, such as cross sections and structure functions, in both the inclusive and diffractive DIS. In CGC calculations, the scattering amplitude of the exclusive diffractive vector meson production in a virtual photon + proton ($\gamma^*+p$) DIS can be factorized into the convolution of three parts, the wave function of virtual photon, $q\bar{q}$ dipole-proton scattering amplitude (dipole amplitude), and vector meson wave function. The key ingredient is the dipole amplitude, since it includes all the QCD dynamics. Especially, it also contains the information about the profile density of the proton. In CGC, one of the widely used dipole amplitudes is the impact parameter dependent saturation (IPsat) model\cite{Kowalski:2003hm}, since the IPsat model contains the information about the impact parameter and gives a rather good description of the experimental data, for instance proton structure function ($F_2$) at HERA\cite{Rezaeian:2012ji}, and charged hadron multiplicity distribution at LHC\cite{Schenke:2013dpa,Tribedy:2011aa}. More specifically, the IPsat model includes the proton profile density function $T_p$, thus it is a considerably suitable model to study the geometric structure of the proton.   

Inspired by the constituent quark picture, a hot spot model was proposed by Mantysaari and Schenke on top of the IPsat model\cite{Mantysaari:2016ykx,Mantysaari:2016jaz}. The hot spot model supposes that the hot spots are inspired by the gluon emission of the respective three constituent quarks, and the transverse positions of these hot spots vary from event to event. The hot spot model provides a rather successful description of the exclusive diffractive vector meson production data at HERA and LHC energies\cite{Mantysaari:2016ykx,Mantysaari:2023xcu,Mantysaari:2022sux,ALICE:2023gcs}, and gives a hint about the lumpy structure of the proton at small $x$. Moreover, it shows that the proton shape fluctuates event-by-event. Thus, it gives an effective initial condition to explain the flow phenomenon in small collision system. Since then, several other models based on the hot spot model were introduced to study the detailed property of the proton\cite{Cepila:2016uku,Kumar:2021zbn,Demirci:2022wuy,Traini:2018hxd,Xiang:2023msj}, see a review\cite{Mantysaari:2020axf} and references therein.    

To explore the possible structure of the proton, we construct a stringy proton model beyond the smallest distance approximation, which is inspired by the quenched lattice QCD calculations in Ref.\cite{Bissey:2006bz}. In our improved stringy proton (ISP) model, the three constituent quarks are connected by gluon tubes which are merged at the Fermat point of the quark triangle. In ISP model, the profile function of the proton is established by summing over the three gluon tubes instead of using the tripe of a typical gluon tube which has the smallest distance (among three distances) between the impact parameter and the gluon tubes.  
The transverse profile of the gluon tubes is assumed to have Gaussian distribution with width $\mathrm{B_{t}}$ ($\mathrm{B_{u}}$ and $\mathrm{B_{d}}$), specifically the $\mathrm{B_{u}}$ for width of up quark induced gluon tube and the $\mathrm{B_{d}}$ for width of down quark induced gluon tube. The positions of the constituent quarks also have Gaussian distribution with width $\mathrm{B_{p}}$. We consider that both the profile of gluon tubes and the positions of the constituent quarks vary from event to event, which can be used to study the proton shape fluctuations. From the setups mentioned above, one can know that the proton profile function in the IPsat model is modified by the ISP model. 

The ISP model is used to calculate the $J/\Psi$ production in $\gamma^*+p$ diffractive DIS and compare to the measurements at HERA. We find that the calculations are in good agreement with data. Especially, at small $t$ our results of $J/\Psi$ production in incoherent process are better matching with the data points than the ones from the hot spot model. We would like to note that the authors in Ref.\cite{Mantysaari:2016jaz} also used stringy proton structure to evaluate the differential cross section of $J/\Psi$ production in order to investigate the proton shape fluctuations. However, they used the smallest distance approximation to simplify the proton profile function. As a consequence, the SD-SP model only includes part of gluon tube fluctuations, which renders other source of fluctuations, like saturation momentum ($Q_s$) and dipole size fluctuations, need to be considered in order to get a reasonable description of the HERA data at small $t$\cite{Mantysaari:2016jaz}. Although the smallest distance approximated stringy proton (SD-SP) model does not fully reflect the stringy structure of the proton, the SD-SP model does reach the aim of testing the proton shape fluctuations. Meanwhile, we find that the radius of proton calculated by the width parameters of the ISP model is consistent with the one from fitting to the recent data from the GlueX collaboration at Jefferson Lab\cite{GlueX:2019mkq,Kharzeev:2021qkd}, which solves a tension that the proton radius extracted from the SD-SP model is incompatible with the measured radius.
Moreover, we study the fine structure of the gluon tubes which connect constituent quarks and Fermat point of the quark triangle. We compute the differential cross section of $J/\Psi$ production in two cases, $\mathrm{B_u\geq B_d}$ and the reverse ($\mathrm{B_u<B_d}$), and compare the results to the HERA data. An interesting result is found that it seems the width of the up quark induced gluon tube always larger than the width of the down quark induced gluon tube in the event-by-event cases, which indicates that the up quark may emit more gluons than the down quark in the proton at small $x$. This finding is consistent with our previous outcomes obtained in the hot spot model\cite{Xiang:2023msj}. 

%


\section{Exclusive diffractive vector meson production in CGC}

A brief review about formalism of calculating the exclusive diffractive vector meson production in $\gamma^* + p$ DIS is present in this section. All the calculations are based on the CGC framework and dipole picture. Generally, there are two types of exclusive diffractive processes in terms of the scattered target proton dissociated or not, see Fig.\ref{eddis}. If the target proton keeps intact after the scattering, we call it as the coherent process. While the target proton is dissociated after the interaction, it is known as the incoherent process. The Good-Walker picture is a widely used approach to describe the exclusive diffractive process\cite{Good:1960ba}. In this picture, the differential cross section of coherent process is proportional to the square of the first moment of the diffractive scattering amplitude, and can be written as\cite{Kowalski:2006hc}   
\be
\frac{d\sigma^{\gamma^*+p\rightarrow V+p}}{dt} = \frac{(1+\beta^2)R_g^2}{16\pi}\Big|\big\langle \mathcal{A}^{\gamma^*+p\rightarrow V+p}(x,Q^2,\bm{\Delta})\big\rangle\Big|^2,
\label{cohamp}
\ee
where $\langle\cdots\rangle$ refers to the average over possible configurations of the target proton, $\mathcal{A}^{\gamma^*+p\rightarrow V+p}$ is the diffractive scattering amplitude which will be discussed in detail below, $Q^2$ is the virtuality of the virtual photon, $\bm{\Delta}\equiv \sqrt{-t}$ is the momentum transfer between the incoming and outgoing proton, and $\gamma^*$, $p$ and $V$ denote the virtual photon, proton and vector meson, respectively. In Eq.(\ref{cohamp}), we take into account the real part correction of the diffractive scattering amplitude by multiplying $(1+\beta^2)$, since the diffractive scattering amplitude was derived in the assumption that it is purely imaginary. $\beta$ is the ratio of the real to imaginary parts of the scattering amplitude
\be
\beta = \tan\Big(\frac{\pi\lambda}{2}\Big)
\ee
with
\be
\lambda = \frac{\partial\ln\big(\mathcal{A}^{\gamma^*+p\rightarrow V+p}\big)}{\partial\ln(1/x)}.
\label{lam}
\ee
The factor $R_g$ in Eq.(\ref{cohamp}) is called the skewedness correction\cite{Watt:2007nr},
\be
R_g =\frac{2^{2\lambda+3}}{\sqrt{\pi}}\frac{\Gamma(\lambda+5/2)}{\Gamma(\lambda+4)}
\ee
with $\lambda$ defined in Eq.(\ref{lam}), which is introduced to account for the imbalance momentum fractions $x$ and $x'$ of the two exchanged gluons between the $q\bar{q}$ dipole and proton target, see Fig.\ref{eddis}. 

In Good-Walker picture, the differential cross-section of the incoherent process is defined as the difference between the second moment and first moment squared of the diffractive scattering amplitude, and can be written as
\be
\frac{d\sigma^{\gamma^*+p\rightarrow V+p'}}{dt} = \frac{(1+\beta^2)R_g^2}{16\pi}\Bigg(\Big\langle\big|\mathcal{A}^{\gamma^*+p\rightarrow V+p}(x,Q^2,\bm{\Delta})\big|^2\Big\rangle-\Big|\big\langle \mathcal{A}^{\gamma^*+p\rightarrow V+p}(x,Q^2,\bm{\Delta})\big\rangle\Big|^2\Bigg),
\label{incohamp}
\ee
which is a variance. Thus, it can be used to probe the amount of fluctuations in the target wave function. 

To clearly see the function of coherent and incoherent cross sections in the study of spacial structure of proton, we compare Eq.(\ref{cohamp}) with Eq.(\ref{incohamp}), one can see that the coherent cross section depends on the average over the diffractive scattering amplitude and as such probes the average shape of the proton. While the incoherent cross section relies on the variance of the proton, which leads to it sensitive to the internal structure of the proton. Therefore, these two cross sections make it possible to extract the overall and detailed information about the proton structure.  

Now turn to introduce the diffractive scattering amplitudes in Eq.(\ref{cohamp}) and Eq.(\ref{incohamp}). In the CGC framework, the amplitude of the $\gamma^*+p$ diffractive scattering can be factorized into three subprocesses (see Fig.\ref{eddis}): Firstly, the virtual photon fluctuates into a $q\bar{q}$ dipole, where the virtual photon wave function $\Psi$ can be precisely calculated by perturbative QED; Then, the $q\bar{q}$ dipole scatters off the target proton with dipole-proton cross section $d\sigma^{dip}/d^2\bm{b}$; Finally, the scattered dipole combines into a final state vector meson with wave function $\Psi_V$. The scattering amplitude of the exclusive diffractive vector meson production can be obtained by the convolution of the overlap wave function and dipole cross section\cite{Kowalski:2006hc}
\be
\mathcal{A}^{\gamma^*+p\rightarrow V+p}_{T,L}(x, Q^2, \bm{\Delta}) = i \int d^2\bm{r}\int d^2\bm{b}\int\frac{dz}{4\pi}\big(\Psi^*\Psi_V\big)_{T,L}\exp\Big\{-i\big[\bm{b}-(1-z)\bm{r}\big]\cdot\bm{\Delta}\Big\}
\frac{d\sigma^{dip}}{d^2\bm{b}},
\label{dipamp}
\ee
where the two dimensional vector $\bm{\Delta}$ is the Fourier conjugate to the center-of-mass of the dipole $\bm{b}-(1-z)\bm{r}$ with $\bm{b}$ being the impact parameter of the $q\bar{q}$ dipole with respect to the proton target, $\bm{r}$ is the transverse size of $q\bar{q}$ dipole, the subscript $T$ and $L$ stand for the transverse and longitudinal contributions. The $z$ and $1-z$ in Eq.(\ref{dipamp}) refer to the longitudinal momentum fraction of the quark and antiquark, respectively. The dipole-proton cross section, $d\sigma^{dip}/d^2\bm{b}$, is a key ingredient of the diffractive scattering amplitude, it includes all the QCD dynamics of the interactions and also the structure information of the proton. We will discuss it in detail together with the stringy proton model in the next section. 
The $\big(\Psi^*\Psi_V\big)_{T,L}$ in Eq.(\ref{dipamp}) are the overlap wave function between the photon and the vector meson and can be written as\cite{Kowalski:2006hc} 
\be
\big(\Psi^*\Psi_V\big)_T=\hat{e}_fe\frac{N_c}{\pi z(1-z)}\Big\{m_f^2K_0(\epsilon r)\phi_T(r,z)-[z^2+(1-z)^2]\epsilon K_1(\epsilon r)\partial_r\phi_T(r,z)\Big\},
\label{wfT}
\ee
and
\be
\big(\Psi^*\Psi_V\big)_L=\hat{e}_fe\frac{N_c}{\pi}2Qz(1-z)K_0(\epsilon r)\bigg[M_V\phi_L(r,z)+\delta\frac{m_f^2-\nabla_r^2}{M_Vz(1-z)}\phi_L(r,z)\bigg],
\label{wfL}
\ee
where $e=\sqrt{4\pi \alpha_{em}}$, $\delta=1$, $\hat{e}_f$ is the vector meson effective charge (e.g. $\hat{e}_f=2/3$ for $J/\Psi$), $m_f$ are the mass of a quark with flavor $f$. $M_V$ is the mass of the produced vector meson, $N_c=3$ is the number of colors, $K_0$ and $K_1$ are the modified Bessel functions of the second kind with $\epsilon^2\equiv z(1-z)Q^2+m_f^2$, $\nabla_r^2$ is a differential operator and defined as $\nabla_r^2\equiv(1/2)\partial_r+\partial_r^2$. The $\phi_{T,L}(r,z)$ are the scalar part of the vector meson wave functions, which cannot be precisely calculated and need to be modeled. In this work, we will use the boosted Gaussian scalar wave function, since it has been successfully used to describe the variety of diffractive measurements at HERA energies. In boosted Gaussian formalism, the scalar wave functions are given by\cite{Kowalski:2006hc}
\be
\phi_{T,L}(r,z)=\mathcal{N}_{T,L}z(1-z)\exp\bigg(-\frac{m_f^2\mathcal{R}^2}{8z(1-z)}-\frac{2z(1-z)r^2}{\mathcal{R}^2}+\frac{m_f^2\mathcal{R}^2}{2}\bigg).
\ee  
In this work, we will use $J/\Psi$ production as a candidate signal to probe the stringy structure of the proton, since its mass is larger enough to make the perturbative calculation of its photoproduction cross section reliable, and its mass is relatively small to make it having high statistics in the measurements. The relevant parameters of $J/\Psi$'s scalar wave functions, $\mathcal{N}_{T,L}$, $m_f$, $\mathcal{R}$, and $M_V$, are given in Table.\ref{para_wf}. 
\begin{table}
  \centering
  \begin{tabular}{cccccc}
    \hline\hline
    Meson & $M_V$/GeV & $m_f$/GeV & $\mathcal{N}_T$& $\mathcal{N}_L$& $\mathcal{R}^2$/GeV$^{-2}$ \\ \hline
     $J/\psi$ & 3.097 & 1.4 & 0.578  & 0.575  & 2.3 \\
     \hline\hline
  \end{tabular}
  \caption{ Parameters of the ``boosted Gaussian'' wave function for $J/\Psi$\cite{Kowalski:2006hc}.}
  \label{para_wf}
\end{table}   

\section{Dipole-proton cross section}

The dipole-proton cross section in Eq.(\ref{dipamp}), is related to the forward elastic scattering amplitude $N$. According to the optical theorem, the dipole-proton cross section can be written as 
\be
\frac{d\sigma^{\mathrm{dip}}}{d^2\bm{b}}(\bm{b}, \bm{r}, x) = 2N(\bm{b}, \bm{r}, x).
\ee 
In the CGC framework, the rapidity (or $x$) evolution of the dipole scattering amplitude is governed by the JIMWLK\footnote{The JIMWLK is the abbreviation of Jalilian-Marian, Iancu,
McLerran, Weigert, Leonidov, Kovner.} evolution equation\cite{Balitsky:1995ub,Jalilian-Marian:1997qno,Jalilian-Marian:1997jhx,Iancu:2000hn,Ferreiro:2001qy}, which can be reduced to a closed Balitsky-Kovchegov (BK) equation in the mean field approximation\cite{Balitsky:1995ub,Kovchegov:1999ua}. Both the JIMWLK and BK equations are extremely complicated integral differential equations, it is very difficult to solve them analytically in the full dynamics regions, although there are some analytic solutions in the saturation regime\cite{Levin:1999mw,Mueller:2001fv,Xiang:2008wr,Xiang:2019kre,Cai:2023iza}. Meanwhile, several numerical methods have been developed to solve the leading order (LO) and next-to-leading order (NLO) BK equations, the reasonable solutions are obtained, and they can describe some measurements at HERA and LHC energies, e.g. proton structure functions\cite{Lappi:2015fma,Lappi:2016fmu,Cepila:2018faq,Ducloue:2019jmy}. It is known that all these solutions are independent of impact parameter. However, the impact parameter plays a key role in the study of the proton spatial structure, since the experimental measurable quantity (transverse momentum transfer $t$) is the Fourier conjugate to the impact parameter. Thus, we need to have the impact parameter information in the dipole scattering amplitude. Unfortunately, it is known that the numerical solution of BK equation exhibits a strong Coulomb tail, once the impact parameter is included.  
Based on the reasons mentioned above, we choose to use the impact parameter dependent saturation (IPsat) model to be as the dipole scattering amplitude, which is widely used in the literature and successfully used in describing the data at HERA, and LHC energies\cite{Rezaeian:2013tka,Schenke:2013dpa,Tribedy:2011aa}. In the IPsat model, the dipole scattering amplitude is given by\cite{Kowalski:2003hm}
\be
N(\bm{b}, \bm{r}, x)
=1-\exp\bigg[-\frac{\pi^2\bm{r}^2}{2N_c}\alpha_s(\mu^2)xg(x,\mu^2)T_p(\bm{b})\bigg],
\label{ipsat}
\ee
where the $xg(x,\mu^2)$ is the gluon distribution whose evolution satisfies the DGLAP evolution equation. The $\mu$ in Eq.(\ref{ipsat}) is a scale which relates to the $\bm{r}$ as
\be
\mu^2 = \frac{4}{\bm{r}^2} + \mu_0^2,
\ee
and the gluon distribution $xg(x,\mu^2)$ at the initial scale $\mu_0^2$ is
\be
xg(x,\mu_0^2) = A_g x^{-\lambda_g}(1-x)^{5.6},
\ee
where the parameters $\mu_0$, $A_g$, and $\lambda_g$ are determined by the fit to the inclusive DIS reduced cross section data at HERA\cite{Rezaeian:2012ji}.   

\begin{figure}[t!]
\setlength{\unitlength}{1.5cm}
\begin{center}
\epsfig{file=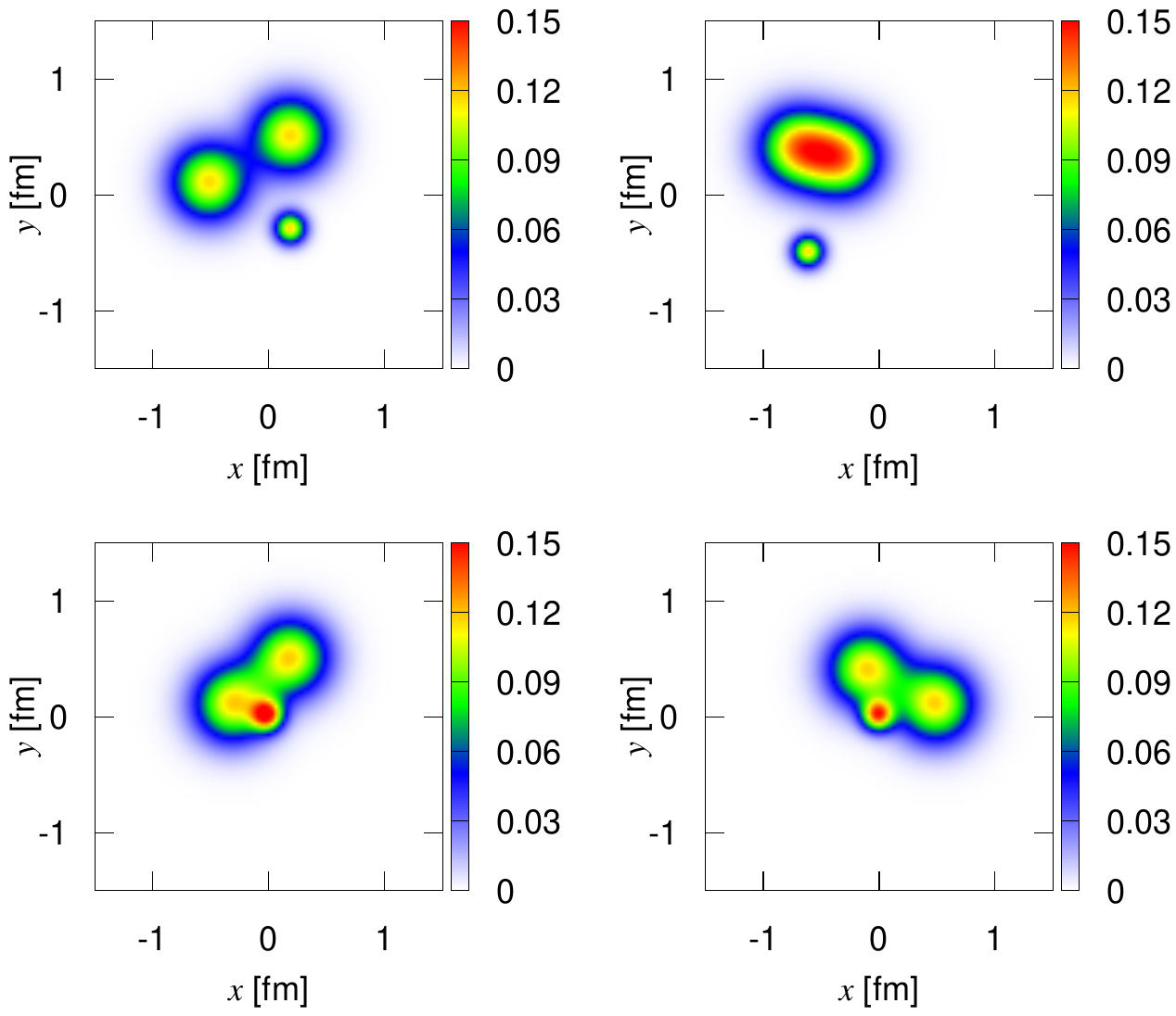, width=7.5cm,height=6.0cm}
\hspace{0.8cm}
\epsfig{file=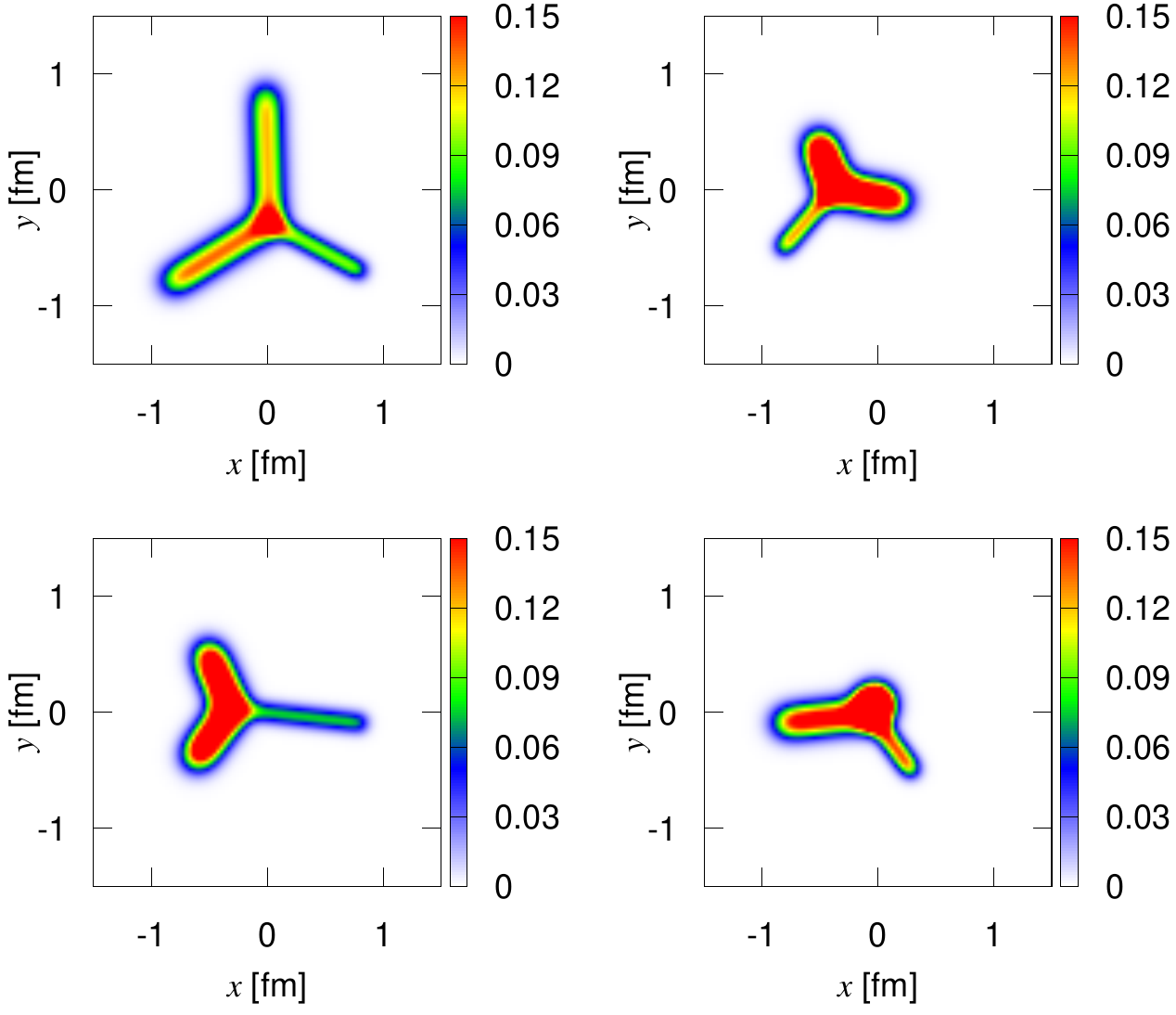, width=7.5cm,height=6.0cm}
\end{center}
\caption{Selected proton density profiles from the hot spot model (left) and stringy proton model (right) at high energy.}
\label{hsgt}
\end{figure}

The transverse profile function, $T_p(\bm{b})$, in Eq.(\ref{ipsat}) includes the spatial structure information of the proton. We use a color string inspired picture to construct the transverse profile function in order to explore the possible structure of the proton. The idea of the stringy proton model is based on the quenched latticed QCD calculations, the three constituent quarks (two up and one down quarks) are connected by the gluon tubes (color fields) which merge at the Fermat point of the quark triangle, see Fig.\ref{hsgt}. We sample the constituent quark positions by a three dimensional Gaussian distribution with width $\mathrm{B_{p}}$. The density profile of the gluon tube, which connects the constituent quarks and Fermat point of the quark triangle, also has a Gaussian distribution and can be obtained by integration over the longitudinal direction\cite{Mantysaari:2016ykx,Mantysaari:2016jaz} 
\be
T_{t}(\bm{b})=\frac{1}{2\pi \mathrm{B_t}}\int db_z\exp\bigg(-\frac{\bm{b}^2+b_z^2}{2\mathrm{B_t}}\bigg),
\label{Tt}
\ee
with width $\mathrm{B_t}$. Then, the profile function of proton, $T_p(\bm{b})$, is established by summing over the three gluon tubes 
\be
T_{p}(\bm{b}) = \frac{1}{N_t}\sum_{i=1}^{N_t}T_t(\bm{b}-\bm{b}_i),
\label{Tp}
\ee
where $N_t=3$ is the number of gluon tubes. We would like to note that a similar stringy proton picture was employed to study the proton shape fluctuations in Ref.\cite{Mantysaari:2016jaz} in which they used the triple of a typical gluon tube which has the smallest distance (among three distances) between the impact parameter and the gluon tubes, to replace the summation of three gluon tubes. As a consequence, the SD-SP model only includes part of gluon tube fluctuations. Therefore, other fluctuations, such as saturation momentum ($Q_s$) and dipole size fluctuations, are needed in order to give a reasonable description of the $J/\Psi$ production data at small $t$. In this paper, we go beyond the smallest distance approximation, an ISP model is established. We find that the ISP model is good enough to reproduce the $J/\Psi$ data at small $t$.  

Let's turn to introduce the $Q_s$ fluctuations. It has been shown that the $Q_s$ fluctuations are significant in the description of the $J/\psi$ production data at small $t$ at HERA energies\cite{Mantysaari:2016ykx,Mantysaari:2016jaz}. We shall consider the $Q_s$ fluctuations followed by Ref.\cite{Mantysaari:2016jaz}, where the saturation scale satisfies a log-normal distribution 
\be
P\big(\ln Q_s^2/\langle Q_s^2\rangle\big)=\frac{1}{\sqrt{2\pi}\sigma}\exp\Bigg[-\frac{\ln^2Q_s^2/\langle Q_s^2\rangle}{2\sigma^2}\Bigg].
\label{lognorm}
\ee
In terms of above distribution, the expectation of $Q_s^2/\langle Q_s^2\rangle$ is
\be
E\big[Q_s^2/\langle Q_s^2\rangle\big]=\exp\big[\sigma^2/2\big].
\label{E_qs}
\ee
If one calculates the $\langle Q_s^2\rangle$, which shows it about $13\%$ (for $\sigma=0.5$) larger than the one without including the $Q_s$ fluctuations. So, one has to normalize the log-normal distribution in order to remain the desired expectation the same. Note that we include the $Q_s$ fluctuations by the way that we let the $Q_s$ of each constituent quark fluctuate independently.

Inspired by our previous findings based on the HS model\cite{Xiang:2023msj}, where we found that the up quark inspired hot spot is different from the down quark inspired hot spot at small $x$. Specifically, each hot spots has different distribution width. In this work, we shall use Eqs.(\ref{Tt}) and (\ref{Tp}) to investigate fine structure of the gluon tube. We let up quark induced two gluon tubes have same width $\mathrm{B_u}$, while the width of down quark induced gluon tube to be $\mathrm{B_d}$. We compare our model calculations with HERA measurements. It shows that the width of the up quark induced gluon tube is always larger than or equal to the width of the down quark induce gluon tube, $\mathrm{B_u\geq B_d}$. This outcome is coincident with the results in Ref.\cite{Xiang:2023msj} where the up quark inspired hot spot width is larger or equal to the down quark inspired hot spot width. These consistent results indicate that the up quark could emit more gluons than the down quark at small $x$.

Besides the stringy proton model mentioned above, we would like to remind that the HS model is a popular candidate to study the spacial structure of the proton in the literature\cite{Cepila:2016uku,Cepila:2023dxn,Kumar:2021zbn,Kumar:2022aly,Kumar:2024kns,Demirci:2022wuy,Traini:2018hxd,Xiang:2023msj}, please see a review and also the reference therein for details of the HS model\cite{Mantysaari:2020axf}. The HS model assumes that the proton consists of several hot spots formed by the gluon emission of constituent quarks (see Fig.\ref{hsgt}), the positions of the hot spots fluctuating from event to event, which leads to the proton shape fluctuates event-by-event. In hot spot model, the profile function of proton is written as\cite{Mantysaari:2016jaz} 
\be
T_{p}(\bm{b}) = \frac{1}{N_q}\sum_{i=1}^{N_q}T_q(\bm{b}-\bm{b}_i),
\ee
where $N_q$ is the number of hot spot, and the profile density of each hot spot is assumed to be Gaussian
\be
T_q(\bm{b}) = \frac{1}{2\pi \mathrm{B_q}}\exp\big[-\frac{\bm{b}^2}{2\mathrm{B_q}}\big]
\ee
with width parameter $\mathrm{B_q}$, where the subscript q denotes u (up) or d (down) quark. The HS model provides reasonable description of the vector meson production at HERA energies, especially for the incoherent measurements. However, the current data at HERA cannot exclude other possible topological structure of the proton due to statistics and limited kinematic regions, e.g. stringy gluon tube configuration. This work shall give an alternative view to know the spacial structure of the proton.     
  
\section{Numerical Results}
\label{res}
In this section, we will give the numerical results about the coherent and incoherent $J/\Psi$ production by using the ISP model. Firstly, we will show comparisons of the $J/\Psi$ differential cross sections between ISP model and HS model. Then, the differential cross sections of $J/\Psi$ production in the cases of $\mathrm{B_u \geq B_d}$ and $\mathrm{B_u < B_d}$ are present. 

\subsection{Comparisons between improved stringy proton and hot spot models}
It is known that the HS model assumes the proton consisting of three hot spots which are formed by the gluon emission of the constituent quarks, and positions of the constituent quarks fluctuating from event to event. The positions of the constituent quarks are sampled randomly in terms of Gaussian distribution. The correlations between constituent quark positions are omitted in the HS model. However, the correlations between the partons could have a significant impact on the proton shape, which is similar to the correlations between the nucleons in a nucleus. As is known that the nucleons' correlation plays an important role on the initial condition of the hydrodynamic evolution of the hot and dense QCD medium. It is difficult to describe the small system flow phenomenon in high energy heavy ion collisions without considering the correlations\cite{Denicol:2014ywa}.  The stringy proton model supposes that the constituent quarks are correlated by gluon tubes (fields). The gluon tubes start from the constituent quarks and end up at the Fermat point of the quark triangle. Thus, in the stringy proton model the spacial structure of the proton is modified by the correlations as compared to the HS model. To clearly see the role of the correlation between the constituent quarks, we calculate the differential cross sections of $J/\Psi$ production by using the HS model (uncorrelated) and ISP model (correlated). We compare the numerical results in order to show the significance of the correlations.      

\begin{figure}[t!]
\setlength{\unitlength}{1.5cm}
\begin{center}
\epsfig{file=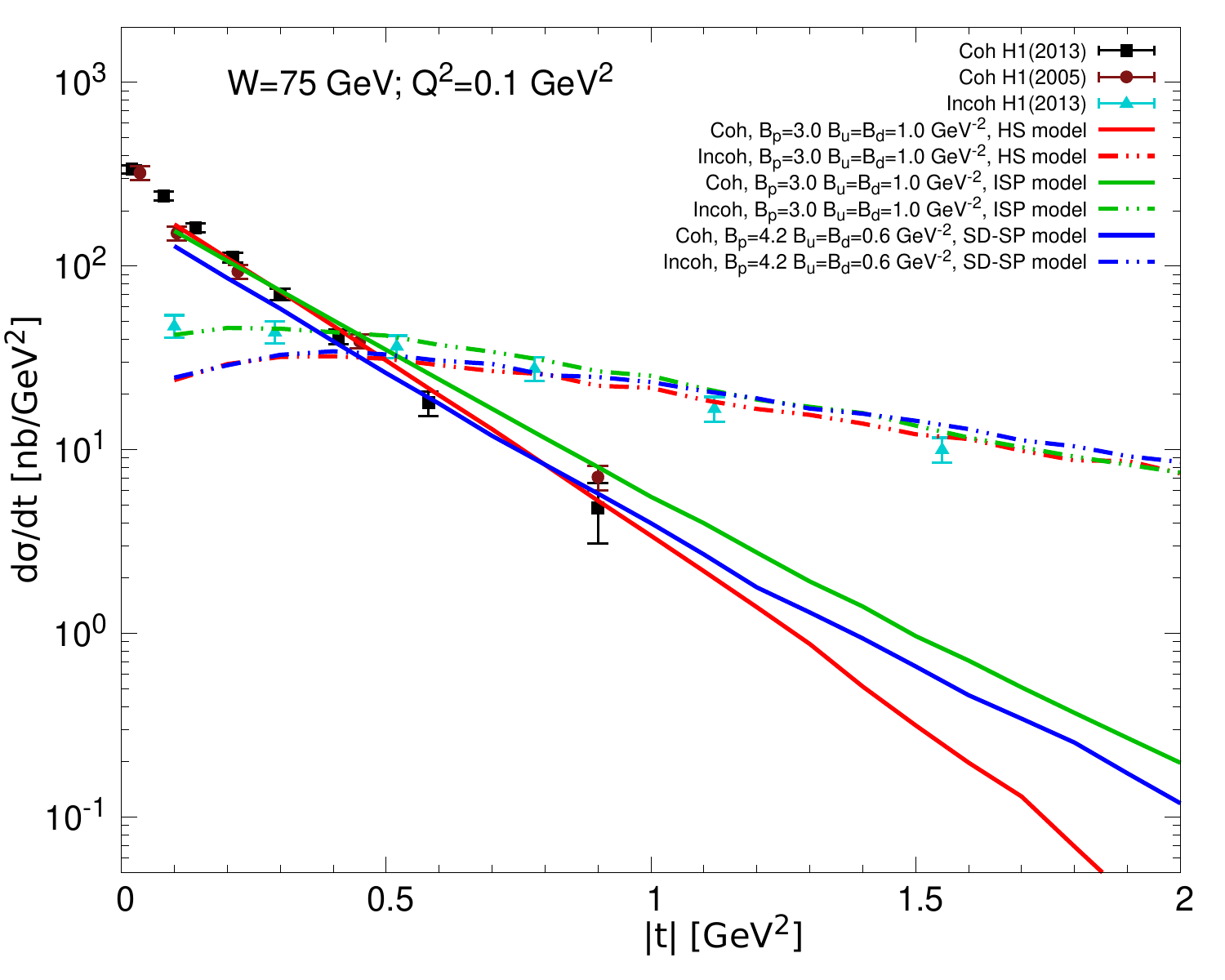, width=8.1cm,height=7.2cm}
\epsfig{file=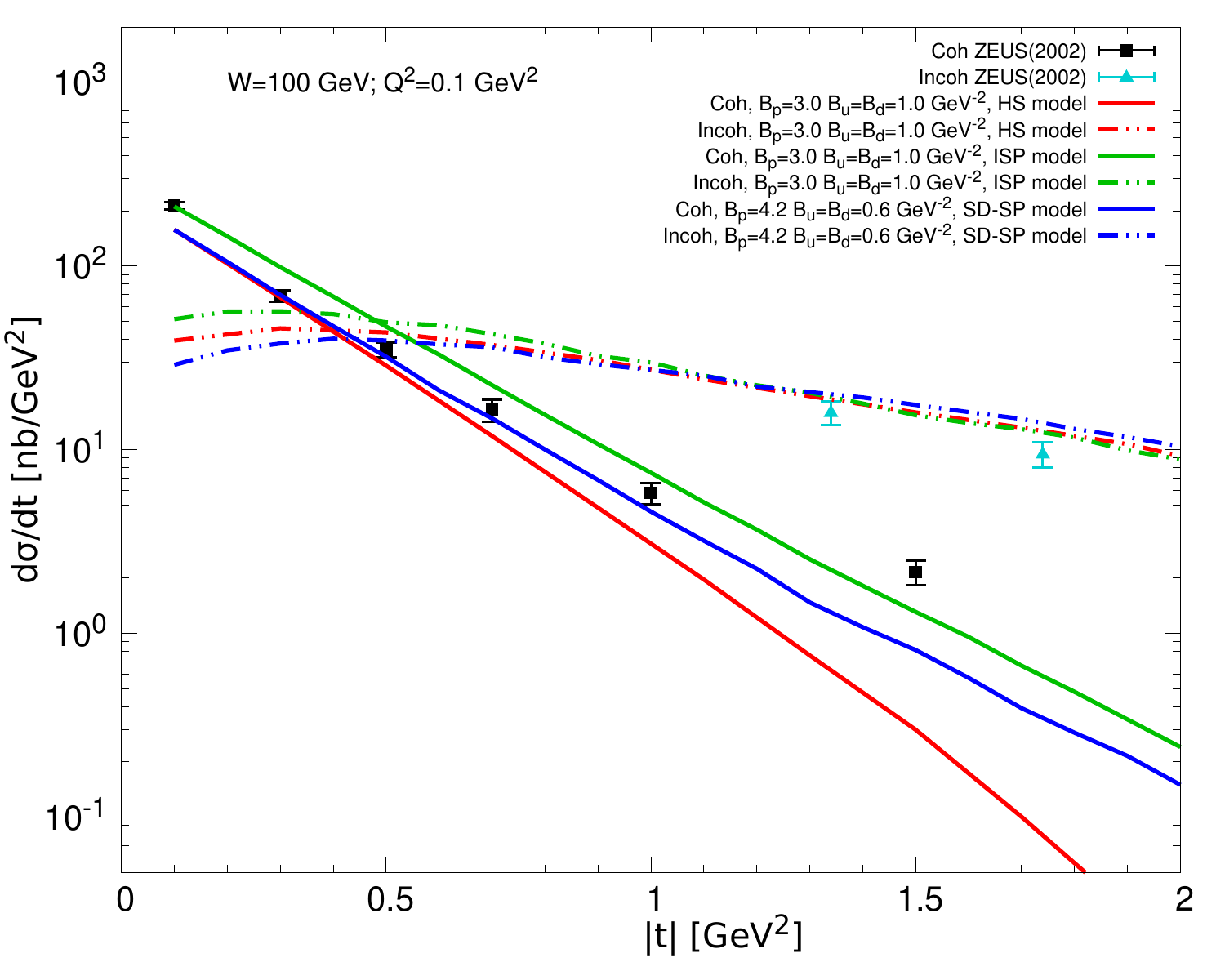, width=8.1cm,height=7.2cm}
\end{center}
\caption{The coherent (solid curves) and incoherent (dashed curves) differential cross section of $J/\Psi$ production as a function of $|t|$ at $\mathrm{W}=75~\mathrm{GeV}$ and $\mathrm{W}=100~\mathrm{GeV}$ as compared to the data from H1 and ZEUS collaborations\cite{H1:2013okq,H1:2005dtp,ZEUS:2002wfj,ZEUS:2002vvv}.}
\label{fig3}
\end{figure}

The coherent and incoherent differential cross sections of the $J/\Psi$ production as a function of transverse momentum transfer $t$ in $\gamma^*+p$ DIS are present in Fig.\ref{fig3}, where the experimental data are taken from H1 and ZEUS Collaborations at HERA\cite{H1:2013okq,H1:2005dtp,ZEUS:2002wfj,ZEUS:2002vvv}. The left panel of Fig.\ref{fig3} gives results calculated at $\mathrm{W}=75~\mathrm{GeV}$, while the right panel of Fig.\ref{fig3} shows the calculations at $\mathrm{W}=100~\mathrm{GeV}$. The sold curves denote the numerical results of the coherent differential cross section, and the dashed curves represent the numerical results of the incoherent differential cross section (similarly hereafter in following figures). Note that the red curves are the results from the HS model, the green curves are the predictions from our ISP model, and the blue curves are the results computed by the SD-SP model. One can see that all the models can give a qualitative description of the coherent and incoherent data at $\mathrm{W}=75~\mathrm{GeV}$ and $\mathrm{W}=100~\mathrm{GeV}$. This outcome seems reasonable due to the fact that all the models include the proton shape fluctuation effect which is a key ingredient in describing the vector meson productions. Although, there is a visible difference of the coherent differential cross sections of $J/\Psi$ production between the HS and ISP models, especially at large $t$, it is hard to see which model is preferred by the data at $\mathrm{W}=75~\mathrm{GeV}$ (see the left panel of Fig.\ref{fig3}), since it only gives the data measured with momentum transfer less than $1~\mathrm{GeV^2}$. Fortunately, the measurements of the coherent $J/\Psi$ production at $\mathrm{W}=100~\mathrm{GeV}$ provide the data up to $t\sim1.5~\mathrm{GeV^2}$, see the right panel of Fig.\ref{fig3}. From the coherent process in the right panel of Fig.\ref{fig3}, one can clearly see that the ISP model is more favored by the data than the HS model at large $t$. 

As is known that the incoherent differential cross section is defined by the variance of the diffractive scattering amplitude, so it is more sensitive to the detailed structure of the proton, namely the details of the model of the proton profile density. We use this feature to show the difference between HS and ISP models in Fig.\ref{fig3}. At a glance, it seems that the dashed curves in Fig.\ref{fig3} successfully match with the data of the incoherent differential cross section of the $J/\Psi$ production in the full $t$ range. However, there are remarkable deviations between the data and calculations from the HS model and SD-SP model at small $t$, although the supplementary saturation scale fluctuations are taken into account. Interestingly, the ISP model gives a rather good description of the incoherent differential cross section data not only at large $t$ but also at small $t$. We trace the reason why our ISP model exhibits a better performance than the other two models, it can attribute to two aspects, correlation and fluctuation. Firstly, The ISP model uses the color string tube to consider the correlations between the constituent quarks, while the HS model ignores the quark correlations. Secondly, the proton profile function in the ISP model is established by summing over the three gluon tubes, in contrast the SD-SP model using the triple of a typical gluon tube which has the smallest distance (among three distances) between the impact parameter and the gluon tubes, to replace the summation of three gluon tubes. This approximation makes the SD-SP model only include part of gluon tube fluctuations, which renders the deviations between the SD-SP model and the data points at small $t$. 

Finally, we would like to mention that the width parameters ($\mathrm{B_p}=3.0~\mathrm{GeV}^{-2}$, $\mathrm{B_u}=1.0~\mathrm{GeV}^{-2}$, and $\mathrm{B_d}=1.0~\mathrm{GeV}^{-2}$) resulting from the ISP model are consistent with ones coming from the HS model\cite{Mantysaari:2016jaz}. When one calculates the root mean square radius of the proton $r_p=\sqrt{2(\mathrm{B_p+B_{u/d}})}$, and gets $r_p=0.55~\mathrm{fm}$, which is in agreement with the one obtained by fitting to the data from the GlueX Collaboration at Jefferson Lab\cite{GlueX:2019mkq,Kharzeev:2021qkd}. However, the proton radius ($r_p=0.61$ fm) resulting from the SD-SP model does not match with the one resulting from fitting to the GlueX data, which implies again that the proton density profile constructed by the sum of the three gluon tubes is more in accordance with the proton geometric structure than the one established by the triple of a typical gluon tube mentioned above.                 

\begin{figure}[t!]
\setlength{\unitlength}{1.5cm}
\begin{center}
\epsfig{file=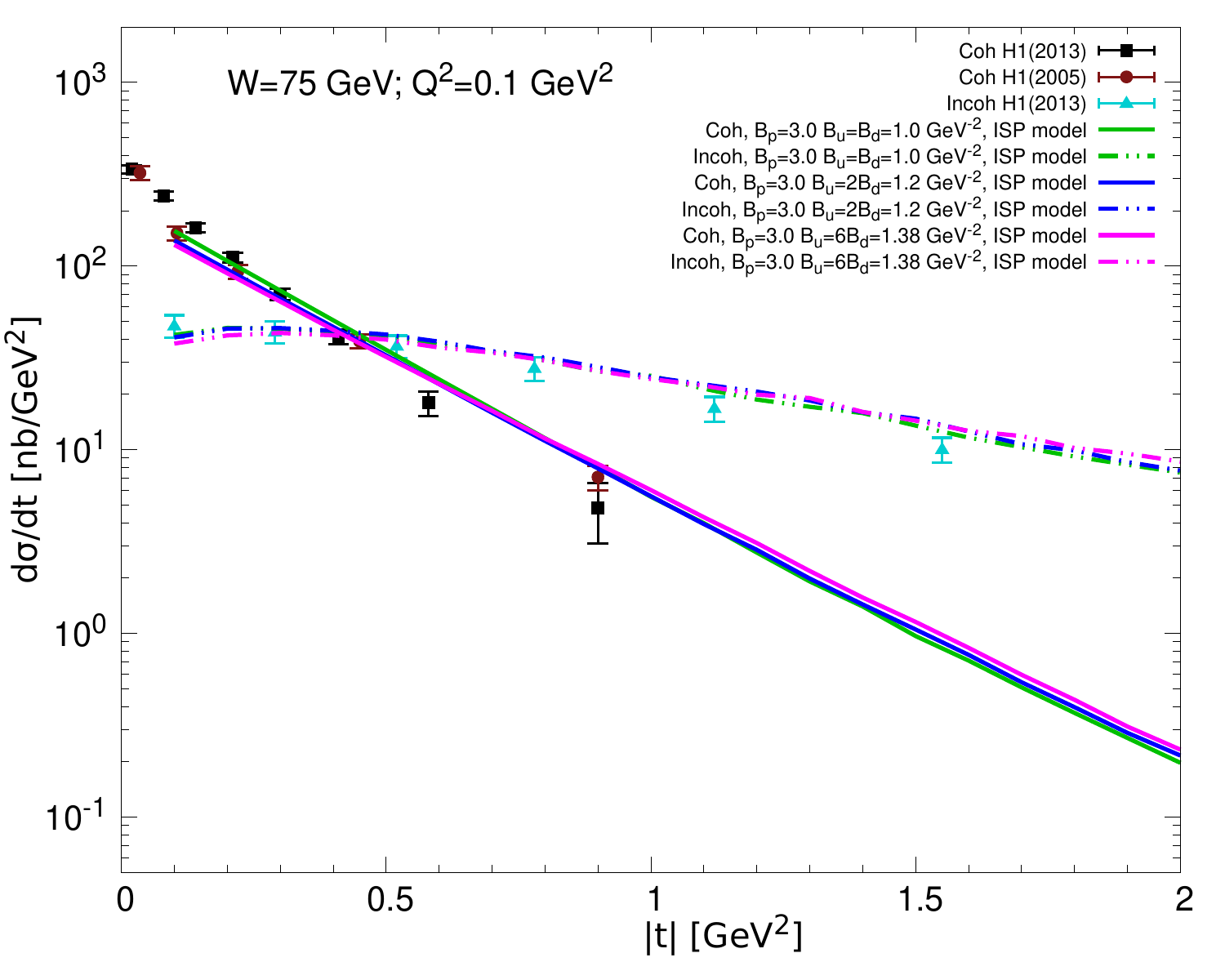, width=8.1cm,height=7.2cm}
\epsfig{file=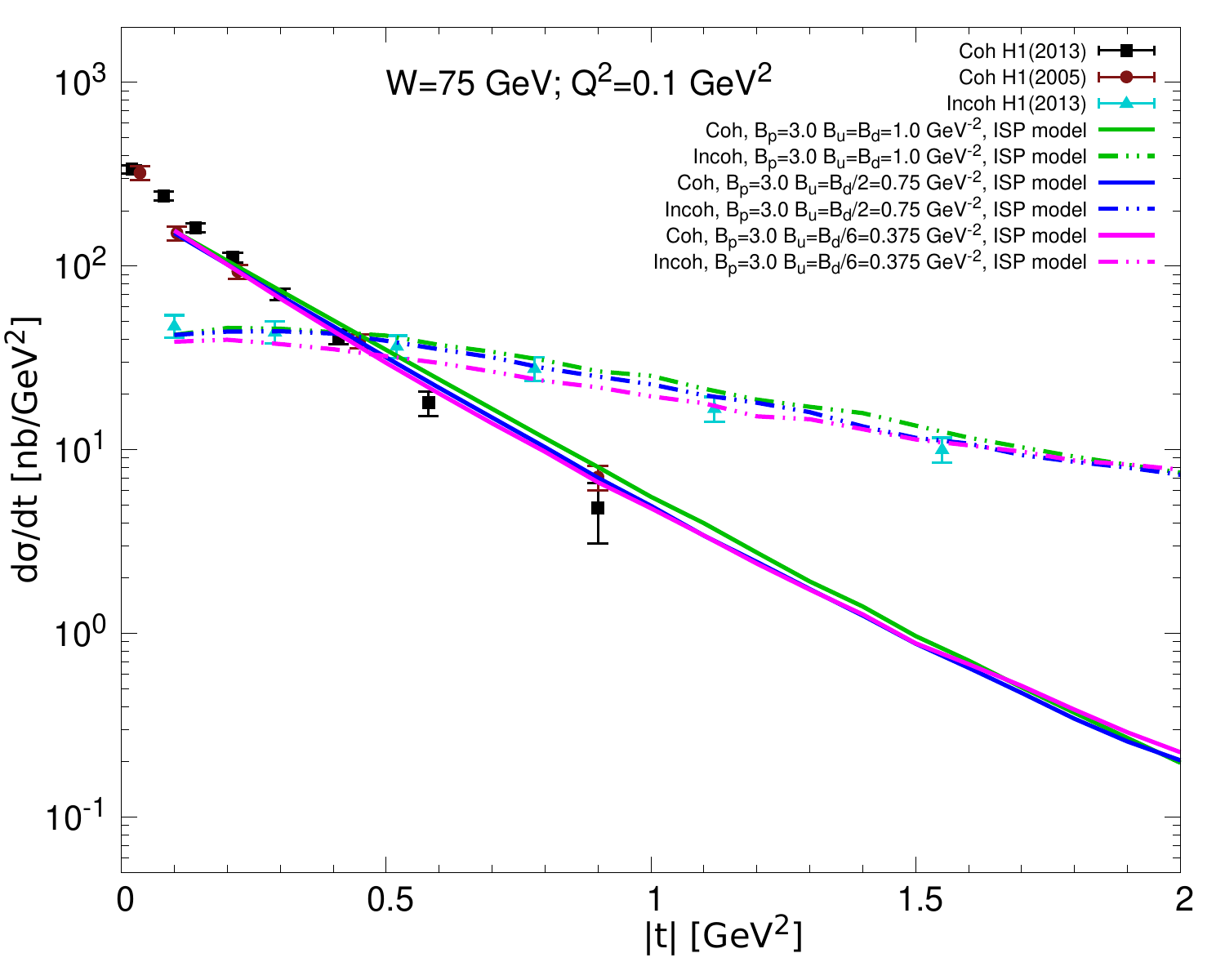, width=8.1cm,height=7.2cm}
\epsfig{file=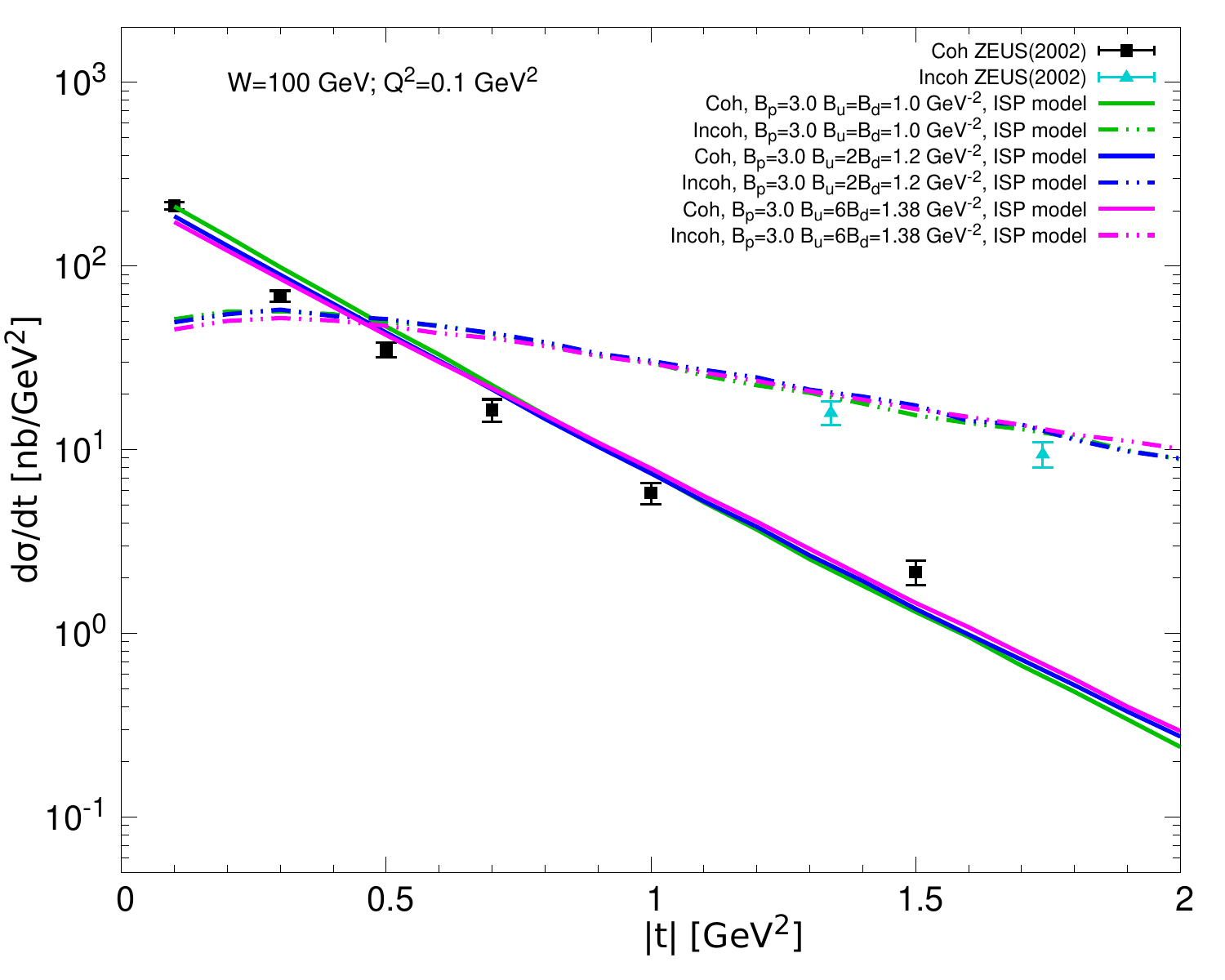, width=8.1cm,height=7.2cm}
\epsfig{file=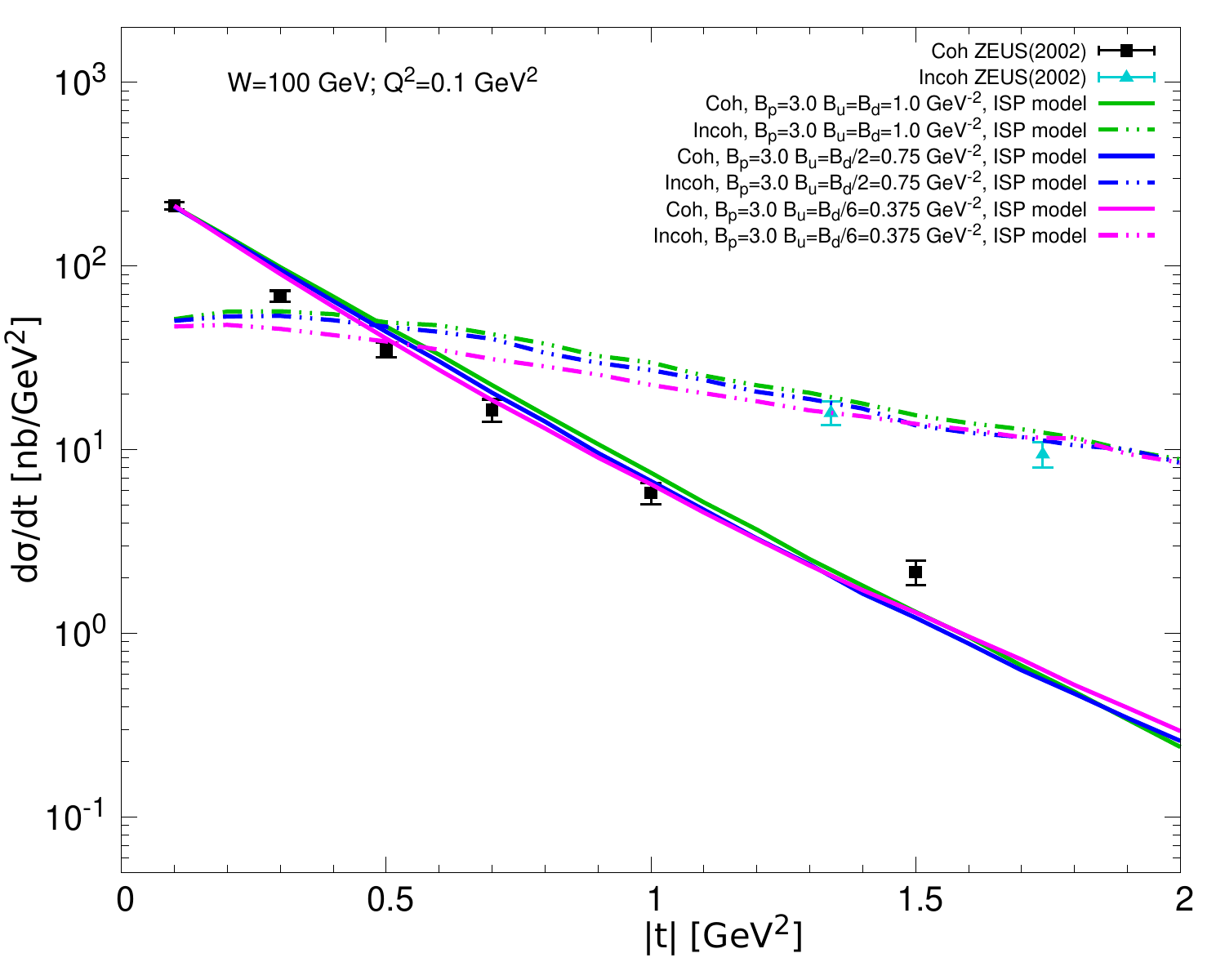, width=8.1cm,height=7.2cm}
\end{center}
\caption{The coherent (solid curves) and incoherent (dashed curves) differential cross section of $J/\Psi$ production as a function of $|t|$ at $\mathrm{W}=75~\mathrm{GeV}$ and $\mathrm{W}=100~\mathrm{GeV}$ as compared to the data from H1 and ZEUS collaborations\cite{H1:2013okq,H1:2005dtp,ZEUS:2002wfj,ZEUS:2002vvv}. The numerical results of left (right) panel are calculated by using $\mathrm{B_u\geq B_d}$ ($\mathrm{B_u< B_d}$). }
\label{fig4}
\end{figure}

\subsection{The fine structure of proton with varied distribution width of the gluon tube}
In recent lattice QCD studies\cite{Bhattacharya:2023ays}, it has been shown that the up quark has a different density distribution from the down quark in the unpolarized proton, and the distortions between the up and down quarks are also different in the polarized proton. These findings indicate that the up quark and down quark in the proton have different structure at small $x$. Moreover, our recent studies in Ref.\cite{Xiang:2023msj} also found that the distribution widths of up and down quarks are different, and it seems that the distribution width of up quark is always larger than or equal to the one of the down quark, which are favored by the HERA data. Inspired by the aforementioned outcomes, we plan to study the fine structure of the gluon tubes which connect the consituent quarks and the Fermat point of the quark triangle (see Fig.\ref{hsgt}), and to see whether the up quark induced gluon tube having different distribution from the down quark induced gluon tube.       

We calculate the coherent and incoherent differential cross sections of the $J/\Psi$ production with varied distribution width of the gluon tube in $\gamma^*+p$ DIS at $\mathrm{W}=75~\mathrm{GeV}$ and $\mathrm{W}=100~\mathrm{GeV}$. The relevant results are shown in Fig.\ref{fig4}. To see the distribution difference between the up and down quarks induced gluon tubes, we vary the gluon tube width parameters $\mathrm{B_{u}}$ and $\mathrm{B_{d}}$, but keeping the average width of gluon tubes $\mathrm{B_{t}}=(2\mathrm{B_{u}}+\mathrm{B_{d}})/3.0=1.0$ unchanged, since it has been shown that $\mathrm{B_t=1.0}$ is an optimal parameter to reproduce the data points\cite{Xiang:2023msj}. Note that in Fig.\ref{fig4} the green curves are calculated by the ISP model with parameters $\mathrm{B_p}=3.0~\mathrm{GeV}^{-2}$ and $\mathrm{B_{u}}$=$\mathrm{B_{d}}$=1.0 $\mathrm{GeV}^{-2}$, which are used to be as the reference to evaluate the compatibility of the other possible distribution widths of the gluon tube, since the green curves give a rather good description of the experimental data. The left (right) panels of Fig.\ref{fig4} demonstrate the numerical results computed by the widths of the up quark induced gluon tubes larger than or equal to (smaller than) the widths of the down quark induced gluon tubes, $\mathrm{B_{u}}\geq \mathrm{B_{d}}$ ($\mathrm{B_{u}} < \mathrm{B_{d}}$). One can see that all the numerical results calculated with $\mathrm{B_{u}}\geq \mathrm{B_{d}}$ are compatible with the HERA data in both coherent and incoherent process, while the predictions computed with $\mathrm{B_{u}} < \mathrm{B_{d}}$ have some tensions with green curves (HERA data). Especially, the deviations are remarkable in the incoherent processes (see the dashed curves on the right panel of Fig.\ref{fig4}), since the incoherent differential cross section is defined by the variance of the diffractive scattering amplitude, it is sensitive to the details of the proton density profile. These outcomes seem to indicate that the up quark could induce more gluon emission than the down quark, which leads to the up quark induced gluon tube having more gluons than the down quark induced gluon tube. These outcomes are consistent with the findings in our previous studies in Ref.\cite{Xiang:2023msj} where we found that the width of the hot spot inspired by gluon emission of the up quark is larger than or equal to the one from the gluon emission of the down quark.

\section{conclusions and discussions}
In order to study the possible geometric structure of the proton, we extend the smallest distance approximated stringy proton model to include the fluctuations from all the three gluon tube contributions. We present a detailed event-by-event calculations of the exclusive diffractive $J/\Psi$ production with our ISP model in the framework of Color Glass Condensate. We find very interesting outcomes that the numerical results of the incoherent differential cross section of $J/\Psi$ production computed by the ISP model at small $t$ are more favored by the HERA data compared to ones resulting from the HS model and the SD-SP model. It shows that the correlation and fluctuation are the reasons why the ISP model has a better performance than the HS model and SD-SP model. Moreover, we find that the width parameters $\mathrm{B_p}$, $\mathrm{B_u}$ and $\mathrm{B_d}$ resulting from the ISP model are more reasonable than the ones from SD-SP model, since the radius of the proton computed by the width parameters of the ISP model is consistent with the one obtained by fitting to the data from the GlueX Collaboration at Jefferson Lab. 

We study the fine structure of the gluon tube within the ISP model. The differential cross sections of exclusive diffractive $J/\Psi$ production are calculated by letting $\mathrm{B_u}$ different from $\mathrm{B_d}$. We find that for the incoherent process all the numerical results of the incoherent differential cross section of $J/\Psi$ production computed with $\mathrm{B_u}\geq \mathrm{B_d}$ are almost consistent with each other at $\mathrm{W}=75 ~\mathrm{GeV}$ and $\mathrm{W}=100~\mathrm{GeV}$, respectively. While the numerical results of the incoherent differential cross section of $J/\Psi$ production calculated with $\mathrm{B_u} < \mathrm{B_d}$ are incompatible with the HERA data. This outcome seems to indicate the possibility that the width of the up quark induced gluon tubes is always larger than or equal to the width of the down quark induced gluon tube, in event-by-event cases. This finding is consistent with the result obtained by the HS model in Ref.\cite{Xiang:2023msj} where the up quark inspired hot spot width can have any times of size larger than or equal to the down quark inspired hot spot width at small $x$.

All the numerical results presented in this paper are calculated by assuming that the positions of the constituent quarks and the profile of the gluon tubes have Gaussian distribution. In fact, they can be other distributions. It has been found if one uses the exponential distribution to be as the position distribution of the constituent quarks, the HS model predictions of the coherent differential cross section of the $J/\Psi$ production are improved at large $t$\cite{Xiang:2024xwc,Goncalves:2024jlx}. The studies of the influence of the position distribution of the constituent quarks and the profile distribution of the gluon tube on the exclusive diffractive $J/\Psi$ production in DIS process will be our next work, which shall provide further insights on the proton structure.      


\begin{acknowledgments}
This work is supported by the National Natural Science Foundation of China under Grant Nos.12165004; the Basic and Applied Basic Research Project of Guangzhou Science and Technology Bureau under Grand No.202201011324; the Education Department of Guizhou Province under Grant No.QJJ[2022]016; the National Key Research and Development Program of China under Grant Nos.2018YFE0104700, and CCNU18ZDPY04.
\end{acknowledgments}

\bibliographystyle{JHEP-2modlong}
\bibliography{refs}

\providecommand{\href}[2]{#2}\begingroup\raggedright\begin{thebibliography}{10}

\bibitem{AbdulKhalek:2021gbh}
R.~Abdul~Khalek {\em et.~al.}, {\it {Science Requirements and Detector Concepts
  for the Electron-Ion Collider: EIC Yellow Report}},
  \href{http://arXiv.org/abs/2103.05419}{{\tt arXiv:2103.05419
  [physics.ins-det]}}.

\bibitem{LHeC:2020van}
{\bf LHeC, FCC-he Study Group} collaboration, P.~Agostini {\em et.~al.}, {\it
  {The Large Hadron-Electron Collider at the HL-LHC}},
  \href{http://dx.doi.org/10.1088/1361-6471/abf3ba}{{\em J. Phys. G} {\bf 48}
  (2021)~no.~11 110501} [\href{http://arXiv.org/abs/2007.14491}{{\tt
  arXiv:2007.14491 [hep-ex]}}].

\bibitem{Anderle:2021wcy}
D.~P. Anderle {\em et.~al.}, {\it {Electron-ion collider in China}},
  \href{http://dx.doi.org/10.1007/s11467-021-1062-0}{{\em Front. Phys.
  (Beijing)} {\bf 16} (2021)~no.~6 64701}
  [\href{http://arXiv.org/abs/2102.09222}{{\tt arXiv:2102.09222 [nucl-ex]}}].

\bibitem{Schlichting:2014ipa}
S.~Schlichting and B.~Schenke, {\it {The shape of the proton at high
  energies}},  \href{http://dx.doi.org/10.1016/j.physletb.2014.10.068}{{\em
  Phys. Lett. B} {\bf 739} (2014) 313}
  [\href{http://arXiv.org/abs/1407.8458}{{\tt arXiv:1407.8458 [hep-ph]}}].

\bibitem{Mantysaari:2016ykx}
H.~M\"antysaari and B.~Schenke, {\it {Evidence of strong proton shape
  fluctuations from incoherent diffraction}},
  \href{http://dx.doi.org/10.1103/PhysRevLett.117.052301}{{\em Phys. Rev.
  Lett.} {\bf 117} (2016)~no.~5 052301}
  [\href{http://arXiv.org/abs/1603.04349}{{\tt arXiv:1603.04349 [hep-ph]}}].

\bibitem{Mantysaari:2016jaz}
H.~M\"antysaari and B.~Schenke, {\it {Revealing proton shape fluctuations with
  incoherent diffraction at high energy}},
  \href{http://dx.doi.org/10.1103/PhysRevD.94.034042}{{\em Phys. Rev. D} {\bf
  94} (2016)~no.~3 034042} [\href{http://arXiv.org/abs/1607.01711}{{\tt
  arXiv:1607.01711 [hep-ph]}}].

\bibitem{Mantysaari:2017cni}
H.~M\"antysaari, B.~Schenke, C.~Shen and P.~Tribedy, {\it {Imprints of
  fluctuating proton shapes on flow in proton-lead collisions at the LHC}},
  \href{http://dx.doi.org/10.1016/j.physletb.2017.07.038}{{\em Phys. Lett. B}
  {\bf 772} (2017) 681} [\href{http://arXiv.org/abs/1705.03177}{{\tt
  arXiv:1705.03177 [nucl-th]}}].

\bibitem{Mantysaari:2022ffw}
H.~M\"antysaari, B.~Schenke, C.~Shen and W.~Zhao, {\it {Bayesian inference of
  the fluctuating proton shape}},
  \href{http://dx.doi.org/10.1016/j.physletb.2022.137348}{{\em Phys. Lett. B}
  {\bf 833} (2022) 137348} [\href{http://arXiv.org/abs/2202.01998}{{\tt
  arXiv:2202.01998 [hep-ph]}}].

\bibitem{Mantysaari:2020lhf}
H.~M\"antysaari, K.~Roy, F.~Salazar and B.~Schenke, {\it {Gluon imaging using
  azimuthal correlations in diffractive scattering at the Electron-Ion
  Collider}},  \href{http://dx.doi.org/10.1103/PhysRevD.103.094026}{{\em Phys.
  Rev. D} {\bf 103} (2021)~no.~9 094026}
  [\href{http://arXiv.org/abs/2011.02464}{{\tt arXiv:2011.02464 [hep-ph]}}].

\bibitem{Cepila:2016uku}
J.~Cepila, J.~G. Contreras and J.~D. Tapia~Takaki, {\it {Energy dependence of
  dissociative $\mathrm{J/}\psi$ photoproduction as a signature of gluon
  saturation at the LHC}},
  \href{http://dx.doi.org/10.1016/j.physletb.2016.12.063}{{\em Phys. Lett. B}
  {\bf 766} (2017) 186} [\href{http://arXiv.org/abs/1608.07559}{{\tt
  arXiv:1608.07559 [hep-ph]}}].

\bibitem{Cepila:2023dxn}
J.~Cepila, J.~G. Contreras, M.~Matas and A.~Ridzikova, {\it {Incoherent
  J/$\psi$ production at large $|t|$ identifies the onset of saturation at the
  LHC}},  \href{http://dx.doi.org/10.1016/j.physletb.2024.138613}{{\em Phys.
  Lett. B} {\bf 852} (2024) 138613}
  [\href{http://arXiv.org/abs/2312.11320}{{\tt arXiv:2312.11320 [hep-ph]}}].

\bibitem{Kumar:2021zbn}
A.~Kumar and T.~Toll, {\it {Investigating the structure of gluon fluctuations
  in the proton with incoherent diffraction at HERA}},
  \href{http://dx.doi.org/10.1140/epjc/s10052-022-10774-3}{{\em Eur. Phys. J.
  C} {\bf 82} (2022)~no.~9 837} [\href{http://arXiv.org/abs/2106.12855}{{\tt
  arXiv:2106.12855 [hep-ph]}}].

\bibitem{Kumar:2022aly}
A.~Kumar and T.~Toll, {\it {Energy dependence of the proton geometry in
  exclusive vector meson production}},
  \href{http://dx.doi.org/10.1103/PhysRevD.105.114011}{{\em Phys. Rev. D} {\bf
  105} (2022)~no.~11 114011} [\href{http://arXiv.org/abs/2202.06631}{{\tt
  arXiv:2202.06631 [hep-ph]}}].

\bibitem{Kumar:2024kns}
A.~Kumar and T.~Toll, {\it {Saturation and fluctuations in the proton
  wavefunction at large momentum transfers in exclusive diffraction at HERA}},
  \href{http://arXiv.org/abs/2403.13631}{{\tt arXiv:2403.13631 [hep-ph]}}.

\bibitem{Xiang:2023msj}
W.~Xiang, Y.~Cai and D.~Zhou, {\it {Imaging constituent quark shape of proton
  with exclusive vector meson production at HERA}},
  \href{http://dx.doi.org/10.1016/j.nuclphysa.2023.122810}{{\em Nucl. Phys. A}
  {\bf 1042} (2024) 122810} [\href{http://arXiv.org/abs/2308.10136}{{\tt
  arXiv:2308.10136 [hep-ph]}}].

\bibitem{Demirci:2022wuy}
S.~Demirci, T.~Lappi and S.~Schlichting, {\it {Proton hot spots and exclusive
  vector meson production}},
  \href{http://dx.doi.org/10.1103/PhysRevD.106.074025}{{\em Phys. Rev. D} {\bf
  106} (2022)~no.~7 074025} [\href{http://arXiv.org/abs/2206.05207}{{\tt
  arXiv:2206.05207 [hep-ph]}}].

\bibitem{Traini:2018hxd}
M.~C. Traini and J.-P. Blaizot, {\it {Diffractive incoherent vector meson
  production off protons: a quark model approach to gluon fluctuation
  effects}},  \href{http://dx.doi.org/10.1140/epjc/s10052-019-6826-0}{{\em Eur.
  Phys. J. C} {\bf 79} (2019)~no.~4 327}
  [\href{http://arXiv.org/abs/1804.06110}{{\tt arXiv:1804.06110 [hep-ph]}}].

\bibitem{Kowalski:2003hm}
H.~Kowalski and D.~Teaney, {\it {An Impact parameter dipole saturation model}},
   \href{http://dx.doi.org/10.1103/PhysRevD.68.114005}{{\em Phys. Rev. D} {\bf
  68} (2003) 114005} [\href{http://arXiv.org/abs/hep-ph/0304189}{{\tt
  arXiv:hep-ph/0304189}}].

\bibitem{Rezaeian:2012ji}
A.~H. Rezaeian, M.~Siddikov, M.~Van~de Klundert and R.~Venugopalan, {\it
  {Analysis of combined HERA data in the Impact-Parameter dependent Saturation
  model}},  \href{http://dx.doi.org/10.1103/PhysRevD.87.034002}{{\em Phys. Rev.
  D} {\bf 87} (2013)~no.~3 034002} [\href{http://arXiv.org/abs/1212.2974}{{\tt
  arXiv:1212.2974 [hep-ph]}}].

\bibitem{Schenke:2013dpa}
B.~Schenke, P.~Tribedy and R.~Venugopalan, {\it {Multiplicity distributions in
  p+p, p+A and A+A collisions from Yang-Mills dynamics}},
  \href{http://dx.doi.org/10.1103/PhysRevC.89.024901}{{\em Phys. Rev. C} {\bf
  89} (2014)~no.~2 024901} [\href{http://arXiv.org/abs/1311.3636}{{\tt
  arXiv:1311.3636 [hep-ph]}}].

\bibitem{Tribedy:2011aa}
P.~Tribedy and R.~Venugopalan, {\it {QCD saturation at the LHC: Comparisons of
  models to p + p and A + A data and predictions for p + Pb collisions}},
  \href{http://dx.doi.org/10.1016/j.physletb.2012.02.047}{{\em Phys. Lett. B}
  {\bf 710} (2012) 125} [\href{http://arXiv.org/abs/1112.2445}{{\tt
  arXiv:1112.2445 [hep-ph]}}].
\newblock [Erratum: Phys.Lett.B 718, 1154--1154 (2013)].

\bibitem{Mantysaari:2023xcu}
H.~M\"antysaari, F.~Salazar and B.~Schenke, {\it {Energy dependent nuclear
  suppression from gluon saturation in exclusive vector meson production}},
  \href{http://dx.doi.org/10.1103/PhysRevD.109.L071504}{{\em Phys. Rev. D} {\bf
  109} (2024)~no.~7 L071504} [\href{http://arXiv.org/abs/2312.04194}{{\tt
  arXiv:2312.04194 [hep-ph]}}].

\bibitem{Mantysaari:2022sux}
H.~M\"antysaari, F.~Salazar and B.~Schenke, {\it {Nuclear geometry at high
  energy from exclusive vector meson production}},
  \href{http://dx.doi.org/10.1103/PhysRevD.106.074019}{{\em Phys. Rev. D} {\bf
  106} (2022)~no.~7 074019} [\href{http://arXiv.org/abs/2207.03712}{{\tt
  arXiv:2207.03712 [hep-ph]}}].

\bibitem{ALICE:2023gcs}
{\bf ALICE} collaboration, S.~Acharya {\em et.~al.}, {\it {First Measurement of
  the |t| Dependence of Incoherent J/\ensuremath{\psi} Photonuclear
  Production}},  \href{http://dx.doi.org/10.1103/PhysRevLett.132.162302}{{\em
  Phys. Rev. Lett.} {\bf 132} (2024)~no.~16 162302}
  [\href{http://arXiv.org/abs/2305.06169}{{\tt arXiv:2305.06169 [nucl-ex]}}].

\bibitem{Mantysaari:2020axf}
H.~M\"antysaari, {\it {Review of proton and nuclear shape fluctuations at high
  energy}},  \href{http://dx.doi.org/10.1088/1361-6633/aba347}{{\em Rept. Prog.
  Phys.} {\bf 83} (2020)~no.~8 082201}
  [\href{http://arXiv.org/abs/2001.10705}{{\tt arXiv:2001.10705 [hep-ph]}}].

\bibitem{Bissey:2006bz}
F.~Bissey, F.-G. Cao, A.~R. Kitson, A.~I. Signal, D.~B. Leinweber, B.~G.
  Lasscock and A.~G. Williams, {\it {Gluon flux-tube distribution and linear
  confinement in baryons}},
  \href{http://dx.doi.org/10.1103/PhysRevD.76.114512}{{\em Phys. Rev. D} {\bf
  76} (2007) 114512} [\href{http://arXiv.org/abs/hep-lat/0606016}{{\tt
  arXiv:hep-lat/0606016}}].

\bibitem{GlueX:2019mkq}
{\bf GlueX} collaboration, A.~Ali {\em et.~al.}, {\it {First Measurement of
  Near-Threshold J/\ensuremath{\psi} Exclusive Photoproduction off the
  Proton}},  \href{http://dx.doi.org/10.1103/PhysRevLett.123.072001}{{\em Phys.
  Rev. Lett.} {\bf 123} (2019)~no.~7 072001}
  [\href{http://arXiv.org/abs/1905.10811}{{\tt arXiv:1905.10811 [nucl-ex]}}].

\bibitem{Kharzeev:2021qkd}
D.~E. Kharzeev, {\it {Mass radius of the proton}},
  \href{http://dx.doi.org/10.1103/PhysRevD.104.054015}{{\em Phys. Rev. D} {\bf
  104} (2021)~no.~5 054015} [\href{http://arXiv.org/abs/2102.00110}{{\tt
  arXiv:2102.00110 [hep-ph]}}].

\bibitem{Good:1960ba}
M.~L. Good and W.~D. Walker, {\it {Diffraction disssociation of beam
  particles}},  \href{http://dx.doi.org/10.1103/PhysRev.120.1857}{{\em Phys.
  Rev.} {\bf 120} (1960) 1857}.

\bibitem{Kowalski:2006hc}
H.~Kowalski, L.~Motyka and G.~Watt, {\it {Exclusive diffractive processes at
  HERA within the dipole picture}},
  \href{http://dx.doi.org/10.1103/PhysRevD.74.074016}{{\em Phys. Rev. D} {\bf
  74} (2006) 074016} [\href{http://arXiv.org/abs/hep-ph/0606272}{{\tt
  arXiv:hep-ph/0606272}}].

\bibitem{Watt:2007nr}
G.~Watt and H.~Kowalski, {\it {Impact parameter dependent colour glass
  condensate dipole model}},
  \href{http://dx.doi.org/10.1103/PhysRevD.78.014016}{{\em Phys. Rev. D} {\bf
  78} (2008) 014016} [\href{http://arXiv.org/abs/0712.2670}{{\tt
  arXiv:0712.2670 [hep-ph]}}].

\bibitem{Balitsky:1995ub}
I.~Balitsky, {\it {Operator expansion for high-energy scattering}},
  \href{http://dx.doi.org/10.1016/0550-3213(95)00638-9}{{\em Nucl. Phys. B}
  {\bf 463} (1996) 99} [\href{http://arXiv.org/abs/hep-ph/9509348}{{\tt
  arXiv:hep-ph/9509348}}].

\bibitem{Jalilian-Marian:1997qno}
J.~Jalilian-Marian, A.~Kovner, A.~Leonidov and H.~Weigert, {\it {The BFKL
  equation from the Wilson renormalization group}},
  \href{http://dx.doi.org/10.1016/S0550-3213(97)00440-9}{{\em Nucl. Phys. B}
  {\bf 504} (1997) 415} [\href{http://arXiv.org/abs/hep-ph/9701284}{{\tt
  arXiv:hep-ph/9701284}}].

\bibitem{Jalilian-Marian:1997jhx}
J.~Jalilian-Marian, A.~Kovner, A.~Leonidov and H.~Weigert, {\it {The Wilson
  renormalization group for low x physics: Towards the high density regime}},
  \href{http://dx.doi.org/10.1103/PhysRevD.59.014014}{{\em Phys. Rev. D} {\bf
  59} (1998) 014014} [\href{http://arXiv.org/abs/hep-ph/9706377}{{\tt
  arXiv:hep-ph/9706377}}].

\bibitem{Iancu:2000hn}
E.~Iancu, A.~Leonidov and L.~D. McLerran, {\it {Nonlinear gluon evolution in
  the color glass condensate. 1.}},
  \href{http://dx.doi.org/10.1016/S0375-9474(01)00642-X}{{\em Nucl. Phys. A}
  {\bf 692} (2001) 583} [\href{http://arXiv.org/abs/hep-ph/0011241}{{\tt
  arXiv:hep-ph/0011241}}].

\bibitem{Ferreiro:2001qy}
E.~Ferreiro, E.~Iancu, A.~Leonidov and L.~McLerran, {\it {Nonlinear gluon
  evolution in the color glass condensate. 2.}},
  \href{http://dx.doi.org/10.1016/S0375-9474(01)01329-X}{{\em Nucl. Phys. A}
  {\bf 703} (2002) 489} [\href{http://arXiv.org/abs/hep-ph/0109115}{{\tt
  arXiv:hep-ph/0109115}}].

\bibitem{Kovchegov:1999ua}
Y.~V. Kovchegov, {\it {Unitarization of the BFKL pomeron on a nucleus}},
  \href{http://dx.doi.org/10.1103/PhysRevD.61.074018}{{\em Phys. Rev. D} {\bf
  61} (2000) 074018} [\href{http://arXiv.org/abs/hep-ph/9905214}{{\tt
  arXiv:hep-ph/9905214}}].

\bibitem{Levin:1999mw}
E.~Levin and K.~Tuchin, {\it {Solution to the evolution equation for high
  parton density QCD}},
  \href{http://dx.doi.org/10.1016/S0550-3213(99)00825-1}{{\em Nucl. Phys. B}
  {\bf 573} (2000) 833} [\href{http://arXiv.org/abs/hep-ph/9908317}{{\tt
  arXiv:hep-ph/9908317}}].

\bibitem{Mueller:2001fv}
A.~H. Mueller in {\em {Cargese Summer School on QCD Perspectives on Hot and
  Dense Matter}}, pp.~45--72, 11, 2001.
\newblock \href{http://arXiv.org/abs/hep-ph/0111244}{{\tt
  arXiv:hep-ph/0111244}}.

\bibitem{Xiang:2008wr}
W.~Xiang, {\it {High energy scattering in the saturation regime including
  running coupling and rare fluctuation effects}},
  \href{http://dx.doi.org/10.1103/PhysRevD.79.014012}{{\em Phys. Rev. D} {\bf
  79} (2009) 014012} [\href{http://arXiv.org/abs/0809.2666}{{\tt
  arXiv:0809.2666 [hep-ph]}}].

\bibitem{Xiang:2019kre}
W.~Xiang, Y.~Cai, M.~Wang and D.~Zhou, {\it {High energy asymptotic behavior of
  the $S$-matrix in the saturation region with the smallest dipole running
  coupling prescription}},
  \href{http://dx.doi.org/10.1103/PhysRevD.101.076005}{{\em Phys. Rev. D} {\bf
  101} (2020)~no.~7 076005} [\href{http://arXiv.org/abs/1911.06744}{{\tt
  arXiv:1911.06744 [hep-ph]}}].

\bibitem{Cai:2023iza}
Y.~Cai, X.~Wang and X.~Chen, {\it {Analytic solution of the Balitsky-Kovchegov
  equation with a running coupling constant using the homogeneous balance
  method}},  \href{http://dx.doi.org/10.1103/PhysRevD.108.116024}{{\em Phys.
  Rev. D} {\bf 108} (2023)~no.~11 116024}
  [\href{http://arXiv.org/abs/2311.02672}{{\tt arXiv:2311.02672 [hep-ph]}}].

\bibitem{Lappi:2015fma}
T.~Lappi and H.~M\"antysaari, {\it {Direct numerical solution of the coordinate
  space Balitsky-Kovchegov equation at next to leading order}},
  \href{http://dx.doi.org/10.1103/PhysRevD.91.074016}{{\em Phys. Rev. D} {\bf
  91} (2015)~no.~7 074016} [\href{http://arXiv.org/abs/1502.02400}{{\tt
  arXiv:1502.02400 [hep-ph]}}].

\bibitem{Lappi:2016fmu}
T.~Lappi and H.~M\"antysaari, {\it {Next-to-leading order Balitsky-Kovchegov
  equation with resummation}},
  \href{http://dx.doi.org/10.1103/PhysRevD.93.094004}{{\em Phys. Rev. D} {\bf
  93} (2016)~no.~9 094004} [\href{http://arXiv.org/abs/1601.06598}{{\tt
  arXiv:1601.06598 [hep-ph]}}].

\bibitem{Cepila:2018faq}
J.~Cepila, J.~G. Contreras and M.~Matas, {\it {Collinearly improved kernel
  suppresses Coulomb tails in the impact-parameter dependent Balitsky-Kovchegov
  evolution}},  \href{http://dx.doi.org/10.1103/PhysRevD.99.051502}{{\em Phys.
  Rev. D} {\bf 99} (2019)~no.~5 051502}
  [\href{http://arXiv.org/abs/1812.02548}{{\tt arXiv:1812.02548 [hep-ph]}}].

\bibitem{Ducloue:2019jmy}
B.~Duclou\'e, E.~Iancu, G.~Soyez and D.~N. Triantafyllopoulos, {\it {HERA data
  and collinearly-improved BK dynamics}},
  \href{http://dx.doi.org/10.1016/j.physletb.2020.135305}{{\em Phys. Lett. B}
  {\bf 803} (2020) 135305} [\href{http://arXiv.org/abs/1912.09196}{{\tt
  arXiv:1912.09196 [hep-ph]}}].

\bibitem{Rezaeian:2013tka}
A.~H. Rezaeian and I.~Schmidt, {\it {Impact-parameter dependent Color Glass
  Condensate dipole model and new combined HERA data}},
  \href{http://dx.doi.org/10.1103/PhysRevD.88.074016}{{\em Phys. Rev. D} {\bf
  88} (2013) 074016} [\href{http://arXiv.org/abs/1307.0825}{{\tt
  arXiv:1307.0825 [hep-ph]}}].

\bibitem{Denicol:2014ywa}
G.~S. Denicol, C.~Gale, S.~Jeon, J.~F. Paquet and B.~Schenke, {\it {Effect of
  initial-state nucleon-nucleon correlations on collective flow in
  ultra-central heavy-ion collisions}},
  \href{http://arXiv.org/abs/1406.7792}{{\tt arXiv:1406.7792 [nucl-th]}}.

\bibitem{H1:2013okq}
{\bf H1} collaboration, C.~Alexa {\em et.~al.}, {\it {Elastic and
  Proton-Dissociative Photoproduction of J/psi Mesons at HERA}},
  \href{http://dx.doi.org/10.1140/epjc/s10052-013-2466-y}{{\em Eur. Phys. J. C}
  {\bf 73} (2013)~no.~6 2466} [\href{http://arXiv.org/abs/1304.5162}{{\tt
  arXiv:1304.5162 [hep-ex]}}].

\bibitem{H1:2005dtp}
{\bf H1} collaboration, A.~Aktas {\em et.~al.}, {\it {Elastic J/psi production
  at HERA}},  \href{http://dx.doi.org/10.1140/epjc/s2006-02519-5}{{\em Eur.
  Phys. J. C} {\bf 46} (2006) 585}
  [\href{http://arXiv.org/abs/hep-ex/0510016}{{\tt arXiv:hep-ex/0510016}}].

\bibitem{ZEUS:2002wfj}
{\bf ZEUS} collaboration, S.~Chekanov {\em et.~al.}, {\it {Exclusive
  photoproduction of J / psi mesons at HERA}},
  \href{http://dx.doi.org/10.1007/s10052-002-0953-7}{{\em Eur. Phys. J. C} {\bf
  24} (2002) 345} [\href{http://arXiv.org/abs/hep-ex/0201043}{{\tt
  arXiv:hep-ex/0201043}}].

\bibitem{ZEUS:2002vvv}
{\bf ZEUS} collaboration, S.~Chekanov {\em et.~al.}, {\it {Measurement of
  proton dissociative diffractive photoproduction of vector mesons at large
  momentum transfer at HERA}},
  \href{http://dx.doi.org/10.1140/epjc/s2002-01079-0}{{\em Eur. Phys. J. C}
  {\bf 26} (2003) 389} [\href{http://arXiv.org/abs/hep-ex/0205081}{{\tt
  arXiv:hep-ex/0205081}}].

\bibitem{Bhattacharya:2023ays}
S.~Bhattacharya, K.~Cichy, M.~Constantinou, X.~Gao, A.~Metz, J.~Miller,
  S.~Mukherjee, P.~Petreczky, F.~Steffens and Y.~Zhao, {\it {Moments of proton
  GPDs from the OPE of nonlocal quark bilinears up to NNLO}},
  \href{http://dx.doi.org/10.1103/PhysRevD.108.014507}{{\em Phys. Rev. D} {\bf
  108} (2023)~no.~1 014507} [\href{http://arXiv.org/abs/2305.11117}{{\tt
  arXiv:2305.11117 [hep-lat]}}].

\bibitem{Xiang:2024xwc}
W.~Xiang, Y.~Cai and D.~Zhou, {\it {On possible distribution laws of the
  transverse position and profile density of the constituent quarks in the
  proton at high energies}},  \href{http://arXiv.org/abs/IN PREPERTION}{{\tt
  arXiv:IN PREPERTION}}.

\bibitem{Goncalves:2024jlx}
V.~P. Goncalves, B.~D. Moreira and L.~Santana, {\it {Probing the spatial
  distribution of gluons within the proton in the coherent vector meson
  production at large $|t|$}},  \href{http://arXiv.org/abs/2404.02746}{{\tt
  arXiv:2404.02746 [hep-ph]}}.

\end{thebibliography}\endgroup

\end{document}